\documentclass[english,12pt,preprint]{aastex}
\usepackage[T1]{fontenc}
\usepackage{graphicx}

\usepackage{natbib}
\def\gs{\mathrel{\raise1.16pt\hbox{$>$}\kern-7.0pt %
\lower3.06pt\hbox{{$\scriptstyle \sim$}}}}         %
\def\ls{\mathrel{\raise1.16pt\hbox{$<$}\kern-7.0pt %
\lower3.06pt\hbox{{$\scriptstyle \sim$}}}}         %

\slugcomment{}
\shorttitle{Tracing the Nuclear Accretion History of the Red Galaxy Population}
\shortauthors{}

\usepackage{babel}
\makeatother
\begin{document}

\title{Tracing the Nuclear Accretion History of the Red Galaxy Population}

\author{Kate Brand\altaffilmark{1}, Arjun Dey\altaffilmark{1}, Michael J. I. Brown\altaffilmark{2}, Casey R. Watson\altaffilmark{3}, Buell T. Jannuzi\altaffilmark{1}, Joan R. Najita\altaffilmark{1}, Christopher S. Kochanek\altaffilmark{4}, Joseph C. Shields\altaffilmark{5}, Giovani G. Fazio\altaffilmark{6}, William R. Forman\altaffilmark{6}, Paul J. Green\altaffilmark{6}, Christine J. Jones\altaffilmark{6}, Almus T. Kenter\altaffilmark{6}, Brian R. McNamara\altaffilmark{5}, Steve S. Murray\altaffilmark{6}, Marcia Rieke\altaffilmark{7}, Alexey Vikhlinin\altaffilmark{6}}

\altaffiltext{1}{National Optical Astronomy Observatory, Tucson, AZ 85726-6732; brand@noao.edu} 
\altaffiltext{2}{Princeton University Observatory, Peyton Hall, Princeton, NJ 08544}
\altaffiltext{3}{Department of Physics, The Ohio State University, 174 West 18th 
Avenue, Columbus, OH 43210}
\altaffiltext{4}{Department of Astronomy, The Ohio State University, 140 West 18th 
Avenue, Columbus, OH 43210}
\altaffiltext{5}{Department of Physics and Astronomy, Ohio University, Athens, OH 
45701}
\altaffiltext{6}{Harvard-Smithsonian Center for Astrophysics, 60 Garden Street, 
Cambridge, MA 02138}
\altaffiltext{7}{Steward Observatory, University of Arizona, 933 North Cherry 
Avenue, Tucson, AZ 85721}

\begin{abstract}

We investigate the evolution of the hard X-ray luminosity of the red galaxy population using a large sample of 3316 red galaxies selected over a wide range in redshift ($0.3<z<0.9$) from a 1.4 deg$^2$ region in the Bo\"otes field of the NOAO Deep Wide-Field Survey (NDWFS). The red galaxies are early-type, bulge-dominated galaxies and are selected to have the same evolution corrected, absolute $R$-band magnitude distribution as a function of redshift to ensure we are tracing the evolution in the X-ray properties of a comparable optical population. Using a stacking analysis of 5-ks Chandra/ACIS observations within this field to study the X-ray emission from these red galaxies in three redshift bins, we find that the mean X-ray luminosity increases as a function of redshift. The large mean X-ray luminosity and the hardness of the mean X-ray spectrum suggests that the X-ray emission is largely dominated by AGN rather than stellar sources. The hardness ratio can be reproduced by either an absorbed (${\rm N_H\approx 2 \times 10^{22}\,cm^{-2}}$) $\Gamma$=1.7 power-law source, consistent with that of a population of moderately obscured Seyfert-like AGN, or an unabsorbed $\Gamma$=0.7 source suggesting a radiatively inefficient accretion flow (e.g., an advection-dominated accretion flow).  We also find that the emission from this sample of red galaxies constitutes at least 5\% of the hard X-ray background. These results suggest a global decline in the mean AGN activity of normal early-type galaxies from $z\sim 1$ to the present, which indicates that we are witnessing the tailing off of the accretion activity onto SMBHs in early-type galaxies since the quasar epoch.
\end{abstract}

\keywords{galaxies: active --- galaxies: evolution --- galaxies: elliptical and lenticular, cD --- X-rays --- cosmology: observations}

\section{Introduction}

The red galaxy population hosts the bulk of the stellar mass at low redshift and is predominantly a bulge-dominated population (e.g., \citealt{hog02}; \citealt{kau03}). It is now known that virtually all local bulge-dominated galaxies host a super-massive black hole (SMBH) at their center (e.g., \citealt{beg03}; \citealt{kor95}). These SMBHs are thought to be relics of an earlier accretion phase onto active galactic nuclei (AGN). The recent discovery that the mass of the SMBH is strongly correlated with the velocity dispersion and mass of the galactic bulge (\citealt{mag98}; \citealt{geb00}; \citealt{fer00}) suggests an intimate relationship between the growth of the black hole by accretion processes and the build-up of the galaxy bulge. While a tight correlation exists between the two today, the relationship at earlier times is less clear, with some observational studies favoring near-lockstep evolution \citep{hec04} while others suggest large variations in the relative bulge/black hole growth rates and mass ratio (e.g., \citealt{gru04}). To properly understand this relation, one must have better measurements of the accretion history onto the central SMBH. 

Studies with the Hubble Space Telescope show that the hosts of luminous quasars are generally massive, early type galaxies (e.g., \citealt{mcle01}; \citealt{dun03}). It is thus likely that many red galaxies may host SMBHs which have formed in an earlier epoch of accretion as powerful AGN. It is therefore in this population that one may expect low luminosity and/or obscured AGN to reside.

Hard X-rays provide us with a relatively unbiased measure of the accretion onto SMBHs due to their ability to penetrate all but the most dense columns of obscuring gas. In this paper, we use X-rays to investigate the possibility that a significant amount of accretion onto SMBHs is still occurring within normal early-type galaxies.  We present a preliminary analysis of the accretion history of a population of red galaxies selected from the 1.4 deg$^2$ initial data release of the NOAO Deep Wide-field Survey (NDWFS; \citealt{jan99}; Jannuzi et al. in prep.; Dey et al. in prep.). Our survey demonstrates the power of stacking analyses of X-rays from large, optically selected, samples of galaxies. We assume a $\Lambda$CDM ($\Omega_ {\rm M}=0.3$, $\Omega_ {\Lambda}=0.7$) cosmology and Hubble constant of $H_{0}=70~ {\rm km~s^{-1}Mpc^{-1}}$.

\section{Data}

\subsection{The Red Galaxy Sample}
 
The optical and near-IR emission from red, early-type, bulge-dominated galaxies is dominated by old stellar populations, and the evolution of these systems is well-approximated by ``passively'' evolving models, where the bulk of the stars are formed in a burst at high redshift ($z>2$) and the population subsequently ages gradually with little additional star formation (e.g., \citealt{sta98}; \citealt{koc00}; \citealt{rus03}). The observed optical colors are dominated by a strong 4000 Angstrom break and vary in a relatively well understood manner; modelling therefore yields accurate photometric redshifts (e.g., \citealt{bro03}; \citealt{csa03}). The expected low levels of star formation imply a low contribution to the total X-ray luminosity from sources such as high-mass X-ray binaries which trace the star formation rate (SFR). 

We selected our sample from a 1.4~deg$^2$ area within the NDWFS Bo\"otes field. The NDWFS comprises deep optical and infra-red imaging in two 9 deg$^2$ regions of the sky, one in Bo\"otes and one in Cetus, and the area considered in this paper lies in the northwest corner of the Bo\"otes field. The NDWFS is an ideal survey for identifying large homogeneous populations due to its deep coverage over large areas. 

Our red galaxy sample selection is described in detail by \citet{bro03}. We used Source Extractor, version 2.2.2 \citep{ber96} to detect the galaxies and measure their optical magnitudes. We then estimated photometric redshifts for the sample using a customized routine which simulates the $B_{W}RI$ color-color space using PEGASE2 spectral evolutionary synthesis models \citep{fio97} that assume exponentially declining star formation rates with $e$-folding times between $\tau$ = 0.6 and 15 Gyr (``$\tau$- models''), and a formation redshift of $z_f\approx 4$. Comparison with spectroscopic redshifts demonstrate that the photometric redshift estimates for this sample are reliable out to $z\sim 1$ (they have a 5\% systematic error and a $\pm$5\% 1$\sigma$ uncertainty; \citealt{bro03}). For the red galaxy sample, we chose galaxies with SEDs well-fit by models with $\tau < 4.5$~Gyr, to select objects which map the red envelope of the galaxy color-color distribution and are likely to trace the evolutionary history of modern-day ellipticals. The rest-frame ($z$=0) color of the $\tau = 4.5$~Gyr model with $E(B-V)$ = 0.04 intrinsic dust extinction is $B_{W}-R$ = 1.44. This is consistent with the Sab template of \citet{fuk95}; therefore our $\tau < 4.5$~Gyr criteria roughly selects galaxy types Sab and earlier. 

We estimated the evolution corrected present-day $R$-band absolute magnitude, M$_{\rm R}$, of each of the galaxies. The PEGASE2 $\tau = 1$~Gyr models and observations (\citealt{jor99}; \citealt{sch99}; \citealt{im02}) suggest the magnitude correction is relatively small: $z$=0.9 red galaxies are $\approx$ 0.9 magnitudes brighter in rest-frame $R$-band than $z$=0 red galaxies. We then imposed an absolute magnitude cut M$_{\rm R}<-21.3$ to ensure that the faintest galaxies in the lowest redshift bin were detectable out to a redshift of $z$=0.9, the limit of our sample.

\subsection{X-ray Imaging: The XBo\"otes Survey}

The X-ray data used in this paper are from the Chandra observations of the Bo\"otes field of the NDWFS (Murray et al., submitted; Kenter et al. submitted). A 9 deg$^2$ region of sky, designed to match the area covered with NDWFS Bo\"otes field, was observed in the 0.5-7.0 keV energy range by the Advanced CCD Imaging Spectrometer (ACIS) on the Chandra X-ray Observatory. This comprised 127 pointings of 5 kilosec each. The X-ray number counts and angular correlation function of the detected sources are reported in Kenter et al. (submitted). Brand et al. (in prep.) presents the optical counterparts to the X-ray detected sources. Here we analyze only 1.4 deg$^2$ (roughly 21 ACIS pointings) of the survey area corresponding to the northwest corner of the NDWFS Bo\"otes field, roughly the area covered by the DR2 release of the NDWFS (http://www.noao.edu/noao/noaodeep). 

\section{Stacking Analysis}
\label{sec:stack}

We investigated the evolution in the mean properties of the red galaxies as a function of redshift using a stacking technique (e.g., \citealt{bra01b}; \citealt{nan02}). This technique is based on the assumption that a large sample can provide an ensemble average measurement of the flux from a typical member even if any single galaxy in a given population is undetected by a particular observation. In other words, one can increase the effective observation depth per object by a factor of the sample size: a sample of $10^3$ galaxies effectively increases the depth of our 5 kilosec Chandra exposure to that of a 5~Msec observation on the mean object. In addition, by stacking the X-ray emission from a large sample of optically selected host galaxies, one obtains an averaged signal which will not be affected by AGN variability of individual objects. 

\begin{figure}
\begin{center}
\setlength{\unitlength}{1mm}
\begin{picture}(150,50)
\put(0,-15){\includegraphics{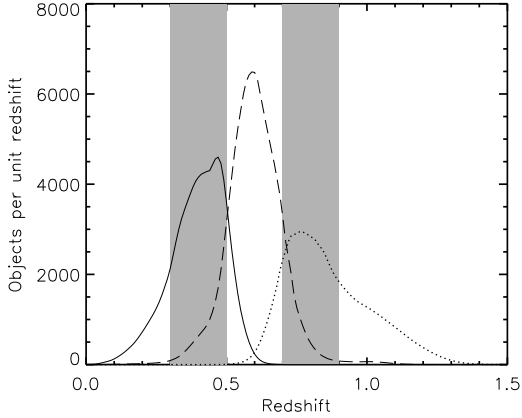}}
\end{picture}
\end{center}
{\caption[junk]{\label{fig:zdist} The sum of the redshift likelihood distributions for all red galaxies within the $0.3<z<0.5$ (solid line), $0.5<z<0.7$ (dashed line), and $0.7<z<0.9$ (dotted line) redshift bins. The redshift distributions of adjoining photometric redshift bins have significant overlaps, but the uncertainty is small enough that the lowest and highest redshift bins should be truly independent.}}
\end{figure}

We divided the sample of 3316 red galaxies into three redshift bins ($0.3<z<0.5$, $0.5<z<0.7$, and $0.7<z<0.9$, consisting of 1040, 1271, and 1005 galaxies respectively) using the photometric redshift estimates described above. Figure~\ref{fig:zdist} shows the sum of the redshift likelihood distributions for the galaxies in each redshift range. Although the errors in the photometric redshifts results in some cross-talk between adjacent redshift bins, they are accurate enough that the lowest and highest redshift bins should be truly independent. The mean present-day rest-frame $R$-band absolute magnitude is constant (M$_{\rm R}$=$-$20.4) for all redshift bins. We created a stacked X-ray image for each redshift bin by centering on each of the optical galaxies in the subsample, effectively extracting the appropriate sub-image from the Chandra image, and summing all the sub-images. 

When considering a stacked signal, it is important to ensure that the measurements are not dominated by individual objects. To check this, we identified the red galaxies within 5 arcsec of an X-ray detection above the total-band (0.5-7.0 keV) false-probability threshold of 5$\times$10$^{-5}$ and $\geq$4 normalized counts (Murray et al. submitted; Kenter et al. submitted). Assuming a standard power-law model spectrum with photon index $\Gamma$=1.7, this detection limit corresponds to luminosities of 3.4 $\times 10^{42}$ ergs s$^{-1}$, 8.6 $\times 10^{42}$ ergs s$^{-1}$, and 16.7 $\times 10^{42}$ ergs s$^{-1}$ in the  $0.3<z<0.5$, $0.5<z<0.7$, and $0.7<z<0.9$ redshift bins respectively. There are 8, 15, and 11 X-ray detected red galaxies, which contribute 45\%, 49\%, and 44\% of the total signal in the $0.3<z<0.5$, $0.5<z<0.7$, and $0.7<z<0.9$ redshift bins. Excluding these sources decreases the signal in approximately the same proportion in all redshift bins, roughly preserving the hardness ratio\footnote{HR = (C$_h$ $-$ C$_s$)/(C$_h$+C$_s$) where C$_h$ and C$_s$ are the counts in the hard (2-7 keV) and soft (0.5-2 keV) bands respectively.} (see Table~\ref{tab:results}). Although our analysis described below excluded the detected sources, our conclusions are only strengthened by their inclusion. 

\begin{figure*}
\begin{center}
\setlength{\unitlength}{1mm}
\begin{picture}(150,60)
\put(-15,-10){\includegraphics{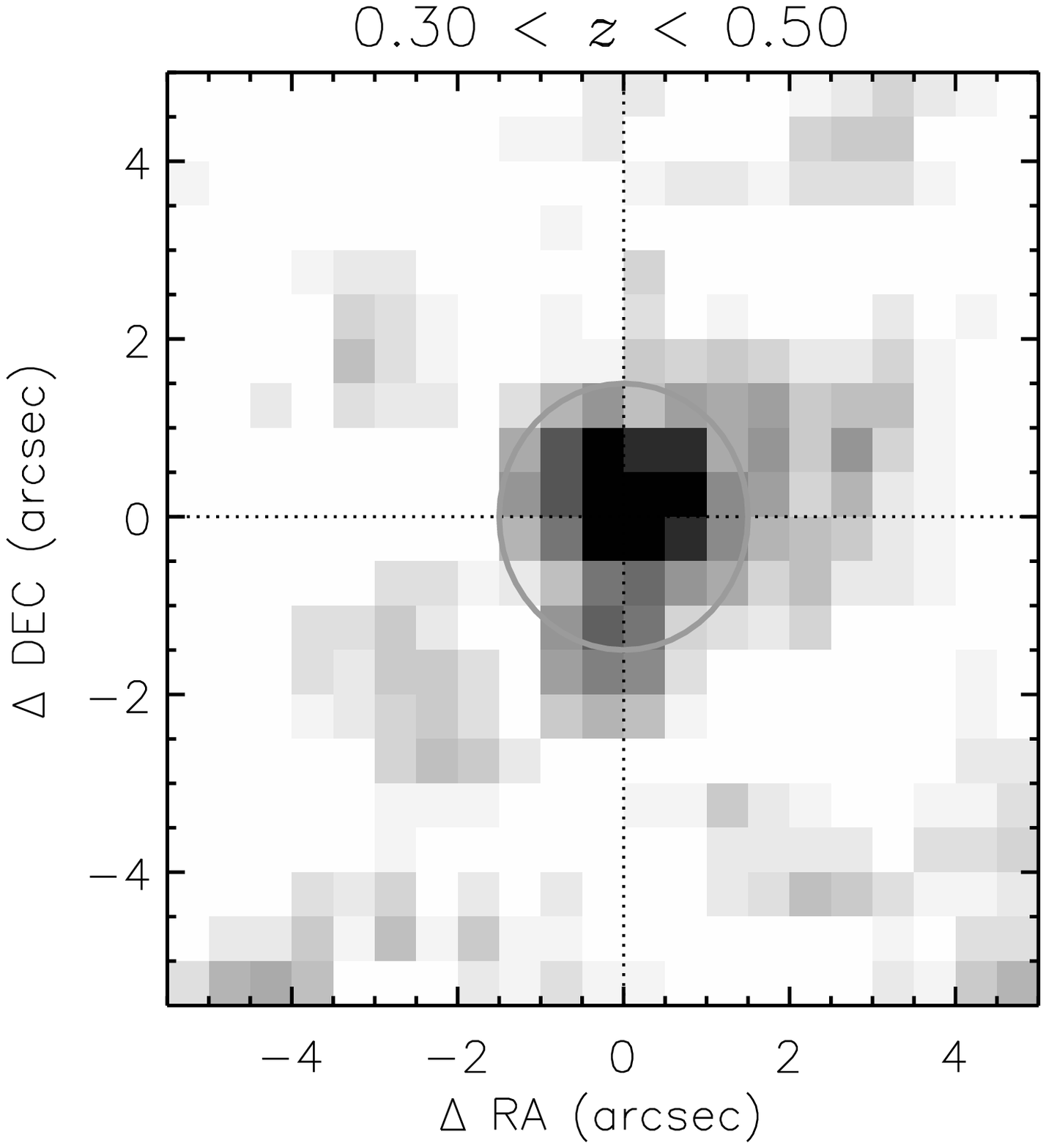}}
\put(40,-10){\includegraphics{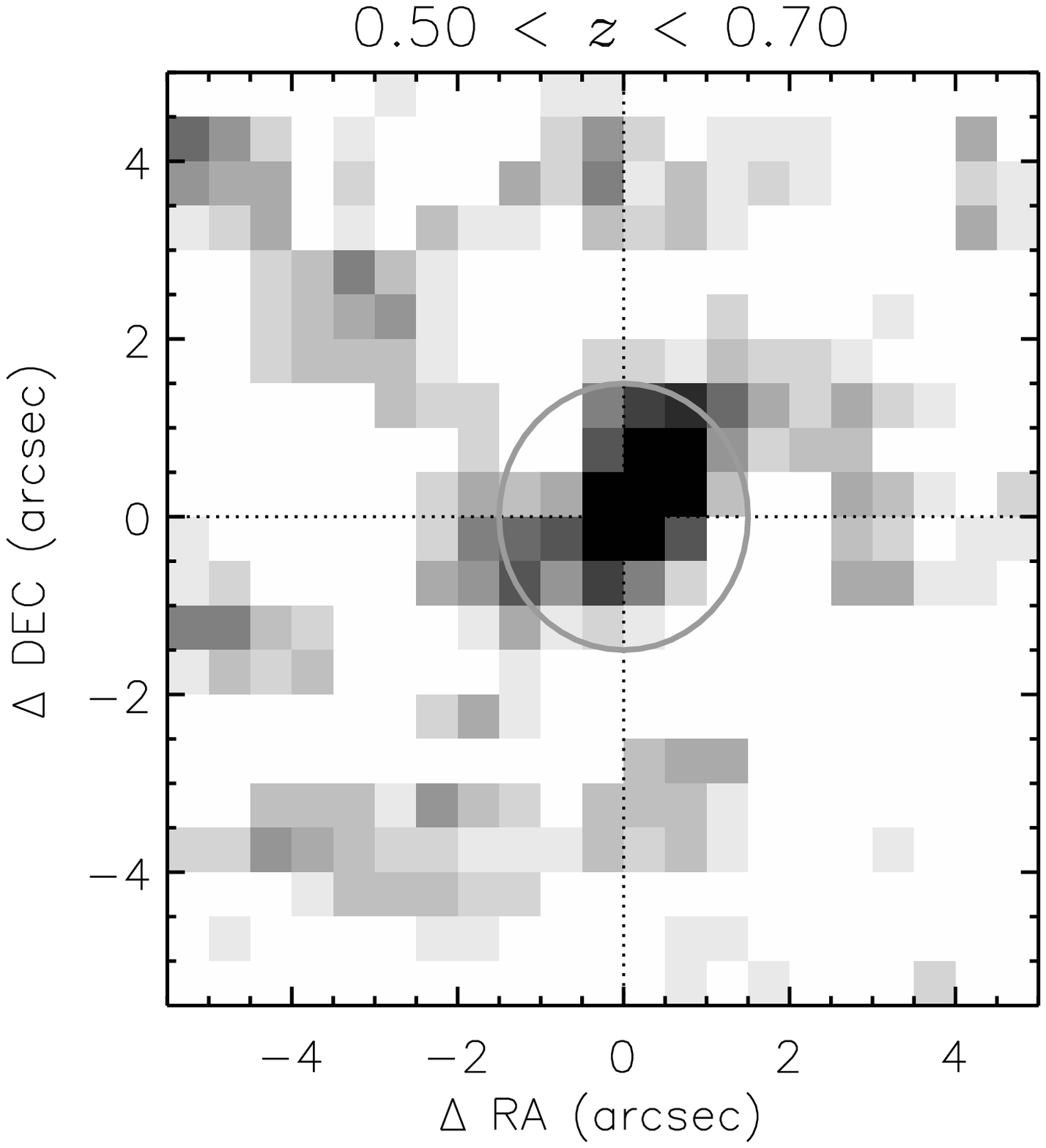}}
\put(95,-10){\includegraphics{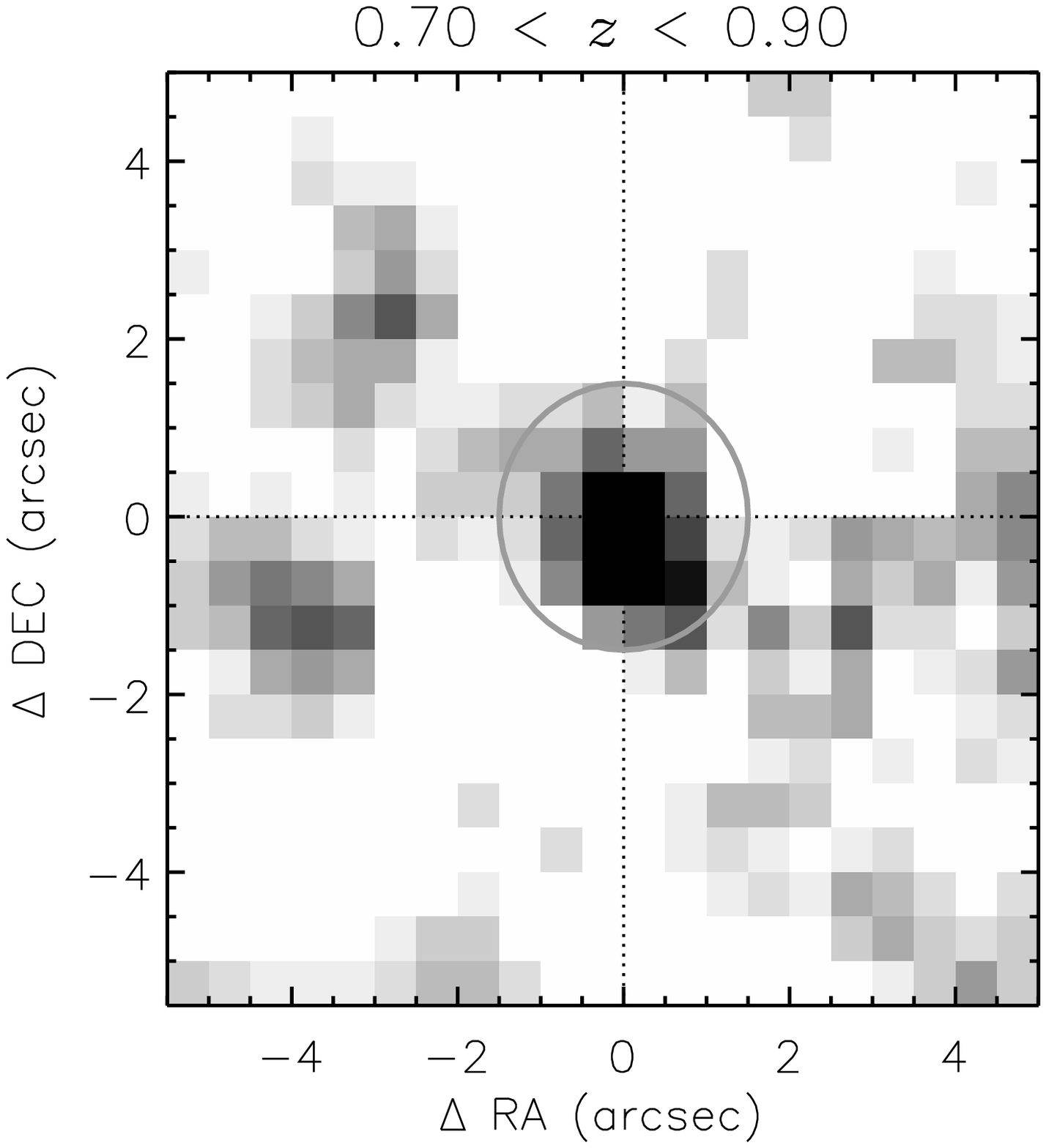}}
\end{picture}
\end{center}
{\caption[junk]{\label{fig:stack_im} The stacked full-band (0.5-7 keV) X-ray images of the individually undetected red galaxies with photometric redshifts 0.3$<z<$0.5 (left), 0.5$<z<$0.7 (center), and 0.7$<z<$0.9 (right). The images are smoothed by a box-car average of 3 pixels. Overplotted is the 1.5 arcsec aperture radius used to extract the signal (see text).}}
\end{figure*}

\begin{figure*}
\begin{center}
\setlength{\unitlength}{1mm}
\begin{picture}(150,140)
\put(-15,-20){\includegraphics{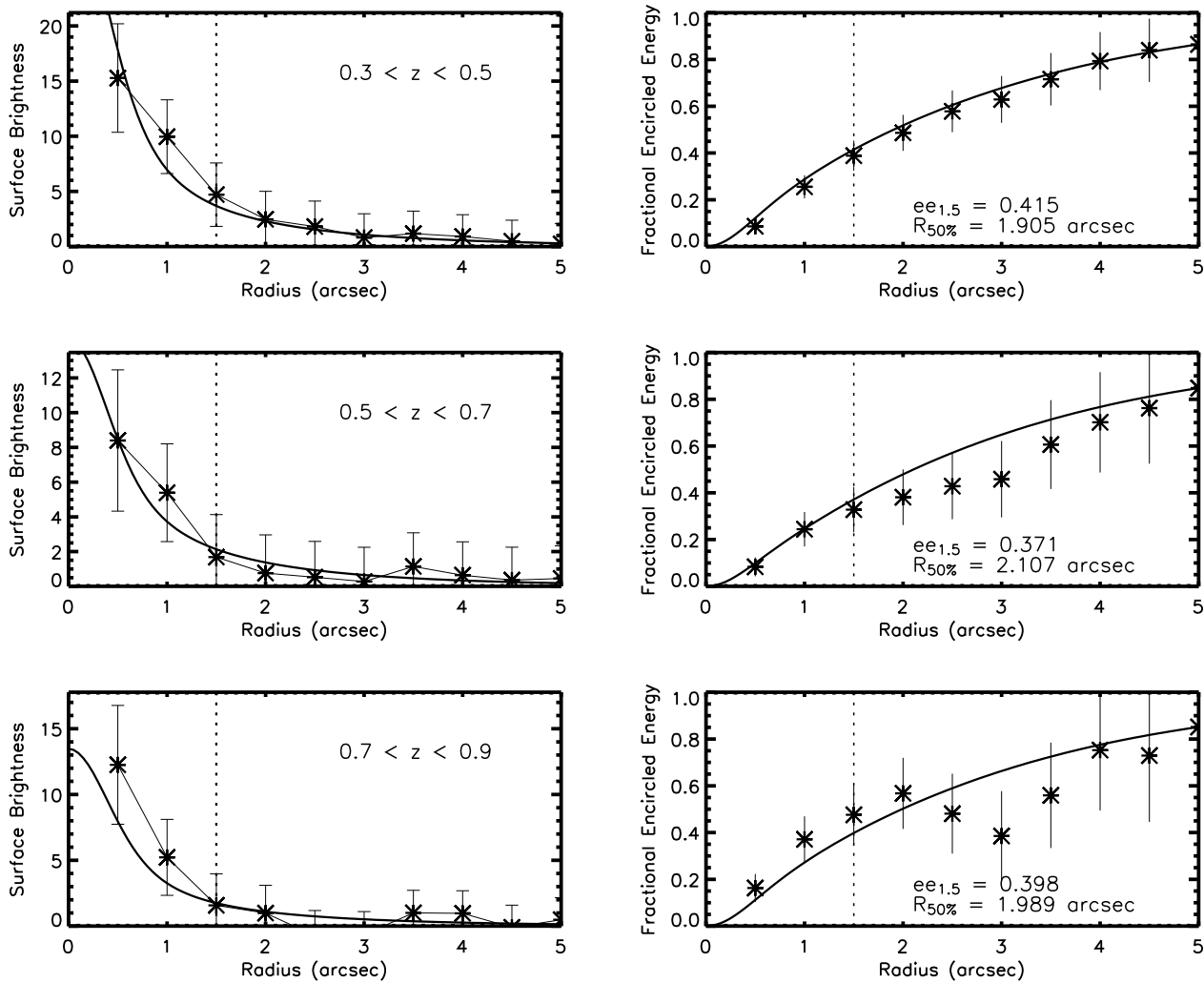}}
\end{picture}
\end{center}
{\caption[junk]{\label{fig:psf_data} Radial surface brightness plots (counts/arcsec$^2$; left) and plots showing the fractional encircled energy as a function of radius (right) for both the data (stars) and integrated PSF models (heavy solid line). These plots are shown for the stacked red galaxies with photometric redshifts 0.3$<z<$0.5 (top), 0.5$<z<$0.7 (middle), and 0.7$<z<$0.9 (bottom). The data was background subtracted using aperture photometry. For the surface brightness profile, the PSF model was normalized to have the same number of total counts as found in the data out to a radius of 5 arcsec. For the fractional encircled energy plots, the data was normalized to have the same fractional encircled energy as the model at 5 arcsec. All errors shown are Poisson counting errors. The vertical dotted line shows the 1.5 arcsec aperture radius in which we performed the aperture photometry.}}
\end{figure*}

After removing all detected sources, Figure~\ref{fig:stack_im} shows that there is a significant detection in all three redshift bins (S/N = 9.7, 4.1, and 5.5 in each of the $0.3<z<0.5$, $0.5<z<0.7$, and $0.7<z<0.9$ redshift bins assuming a 1.5 arcsec aperture radius). Radial profiles of the data are shown in Figure~\ref{fig:psf_data}. Our measurements are presented in Table~\ref{tab:results}. 

\begin{deluxetable}{llll}
\tabletypesize{\scriptsize}
\tablecolumns{4} 
\tablewidth{0pc} 
\tablecaption{\label{tab:results}Table showing the total X-ray counts, the estimated background X-ray counts, the resultant X-ray counts, the detection significance ($\sigma$), and the average X-ray Luminosity in the total (t; 0.5-7 keV), soft (s; 0.5-2 keV), and hard (h; 2-7 keV) bands for each redshift bin. The X-ray luminosity was calculated assuming a power-law spectrum with photon index $\Gamma$=1.7 (PL Lum), a power-law spectrum with photon index $\Gamma$=1.7 and an HI column density of 2$\times$10$^{22}$ cm$^{-2}$ at the mean redshift of the bin (Abs Lum), and a power-law spectrum with photon index $\Gamma$=0.7 ($\Gamma$=0.7 Lum). All models include absorption the Galactic HI column density of 1.75$\times$10$^{20}$ cm$^{-2}$. Also shown is the mean total (0.3-8 keV) band X-ray luminosity predicted for LMXBs (\citealt{kim04}) and the mean hard (2-8 keV) band X-ray luminosity predicted for HMXBs (\citealt{gri03}).} 
\tablehead{ 
\colhead{Redshift} & \colhead{0.3-0.5} & \colhead{0.5-0.7} & \colhead{0.7-0.9} \\
}
\startdata 
\cutinhead{All sources} 
Number of galaxies & 1040 & 1271 & 1005\\
mean $R$ magnitude & 19.75 & 20.95 & 21.85\\
mean M$_{\rm R}$ & $-$21.2 & $-$21.2 & $-$21.2\\
Counts (t/s/h)& 142$\pm$11.9/73$\pm$8.5/69$\pm$8.3 & 110$\pm$10.5/47$\pm$6.9/63$\pm$7.9 & 97$\pm$9.8/52$\pm$7.2/45$\pm$6.7 \\
Bkgd counts (t/s/h)& 26.7$\pm$5.2/10.4$\pm$1.8/15.9$\pm$4.0 & 33.1$\pm$5.8/12.7$\pm$3.6/20.3$\pm$4.5 & 26.6$\pm$5.2/10.3$\pm$3.2/16.0$\pm$4.0\\
Counts-bkgd (t/s/h)& 115.3$\pm$13.0/62.6$\pm$9.1/53.1$\pm$9.2 & 76.9$\pm$10.5/34.4$\pm$7.7/42.7$\pm$9.1 & 70.4$\pm$11.1/41.8$\pm$7.9/29.0$\pm$7.8\\
Detection Significance ($\sigma$)(t/s/h) & 22.3/19.4/13.2 & 13.4/9.7/9.5 & 13.6/13.0/7.3\\
PL Lum ($10^{41}$ erg/s) & 2.3$\pm$0.2/0.7$\pm$0.09/2.4$\pm$0.4 & 3.5$\pm$0.5/0.9$\pm$0.2/4.5$\pm$0.8 & 7.3$\pm$1.0/2.4$\pm$0.4/7.0$\pm$1.6\\
Abs Lum ($10^{41}$ erg/s) & 3.4$\pm$0.3/0.3$\pm$0.04/1.3$\pm$0.2 & 4.8$\pm$0.7/0.4$\pm$0.1/2.1$\pm$0.4 & 9.4$\pm$1.3/1.3$\pm$0.2/3.0$\pm$0.7\\
$\Gamma$=0.7 Lum ($10^{41}$ erg/s) & 2.6$\pm$0.3/0.5$\pm$0.06/2.0$\pm$0.3 & 3.5$\pm$0.5/0.5$\pm$0.1/3.3$\pm$0.6 & 6.6$\pm$0.9/1.3$\pm$0.2/4.5$\pm$1.1\\
\cutinhead{X-ray non-detected sources} 
Number of galaxies & 1032 & 1256 & 994\\
mean $R$ magnitude & 19.75 & 20.95 & 21.85\\
mean M$_{\rm R}$ & $-$21.2 & $-$21.2 & $-$21.2\\
Counts (t/s/h)& 76$\pm$8.7/37$\pm$6.1/39$\pm$6.2 & 56$\pm$7.5/26$\pm$5.1/30$\pm$5.5 & 54$\pm$7.3/23$\pm$4.8/31$\pm$5.6\\
Bkgd counts (t/s/h)& 26.4$\pm$5.1/10.4$\pm$3.2/15.8$\pm$4.0 & 32.7$\pm$5.7/12.5$\pm$3.5/20.0$\pm$4.5 & 26.1$\pm$5.1/10.1$\pm$3.2/15.8$\pm$4.0\\
Counts-bkgd (t/s/h)& 49.6$\pm$10.1/26.7$\pm$6.9/23.2$\pm$7.4 & 23.3$\pm$9.4/13.5$\pm$6.2/10.0$\pm$7.1 & 27.9$\pm$8.9/12.9$\pm$5.8/15.2$\pm$6.8\\
Detection Significance ($\sigma$)(t/s/h) & 9.7/8.3/5.8 & 4.1/3.8/2.2 & 5.5/4.1/3.8\\
PL Lum ($10^{41}$ erg/s) & 1.0$\pm$0.2/0.3$\pm$0.1/1.1$\pm$0.3 & 1.1$\pm$0.3/0.3$\pm$0.1/1.1$\pm$0.6 & 2.9$\pm$0.8/0.8$\pm$0.3/3.7$\pm$1.4\\
Abs Lum ($10^{41}$ erg/s) & 1.5$\pm$0.3/0.1$\pm$0.03/0.6$\pm$0.2 & 1.5$\pm$0.5/0.2$\pm$0.07/0.5$\pm$0.3 & 3.8$\pm$1.0/0.4$\pm$0.2/1.6$\pm$0.6\\
$\Gamma$=0.7 Lum ($10^{41}$ erg/s) & 1.1$\pm$0.2/0.2$\pm$0.05/0.9$\pm$0.2 & 1.1$\pm$0.3/0.2$\pm$0.08/0.8$\pm$0.4 & 2.7$\pm$0.7/0.4$\pm$0.15/2.4$\pm$0.9\\
\hline
\hline
LMXB Lum ($10^{41}$ erg/s) & 0.17$\pm$0.07 & 0.17$\pm$0.07 & 0.18$\pm$0.07 \\
HMXB Lum ($10^{41}$ erg/s) & 0.10 & 0.14 & 0.26 \\
\hline
\enddata
\end{deluxetable} 

We measured the X-ray counts from each stacked image using a centered circular aperture with a fixed angular radius of 1.5 arcsec. We estimated the background by performing aperture photometry on the stacked X-ray image at a angular radius between 10 and 30 arcsec from each red galaxy. After excluding all galaxies that were individually detected in the X-ray, we subtracted our estimate of the background flux from that of the remaining stacked sample. We converted the net flux to an X-ray luminosity using the central redshift of the redshift bin.

Ideally, one would use an aperture radius which has the same physical radius on the sky at different redshifts. This would ensure that the X-ray signal was not increasingly dominated by off-nuclear radiation such as diffuse emission from hot gas and X-ray binaries in more distant sources. However, we chose an aperture of fixed angular size because the effects of losing signal outside the aperture radius due to the Chandra point spread function (PSF) dominate over a contribution from extended emission and a variable size aperture would degrade the signal-to-noise (S/N) at higher redshifts. To justify the choice of a fixed angular aperture radius of 1.5 arcsec, we ran Monte Carlo simulations to ensure any input redshift evolution in luminosity is correctly recovered using this aperture. These tests are described in more detail in Section~\ref{sec:mc}. Performing the analysis with a fixed physical radius does not significantly change our results. 

\subsection{Correcting for the Chandra PSF}

The Chandra point spread function (PSF) broadens markedly as a function of off-axis angle (the 50\% encircled energy radius increases from 0.5 arcsec on-axis to 6.5 arcsec at an off-axis angle of 10 arcmin). This will result in an increasing fraction of the total X-ray counts falling outside the aperture radius for galaxies which happen to lie at a larger off-axis angle from the Chandra pointing center. We corrected for this effect in the following way. Based on a fit to the numerical ray-trace calculations presented in the Chandra proposers' observatory guide, we adopted a model of the Chandra 50\% encircled energy radius, $r_{50}$ at a given off-axis distance, $\Delta\theta$:

\begin{equation}\label{eqn:r50}
r_{50} = 0.43-0.1\Delta\theta+0.05\Delta\theta^2 
\end{equation}

\noindent where $r_{50}$ is in arcsec and $\Delta\theta$ is in arcmin. We then calculated the PSF for each of the red galaxies (each with a different off-axis distance), summed the individual PSFs to create the effective PSF for each stacked image, and then integrated to find the fractional encircled energy as a function of radius. Figure~\ref{fig:psf_data} shows that the shape of the PSF model we used provides a good fit to the data, suggesting that our assumption that the X-ray sources are point sources is a reasonable approximation. The fractional 50\% encircled energy within a 1.5 arcsec radius, $f_{50}$ are 0.415, 0.371, and 0.398 for the $0.3<z<0.5$, $0.5<z<0.7$, and $0.7<z<0.9$ redshift bins respectively (i.e., we detect only approximately 40\% of the total flux due to the effects of the PSF). To account for the losses due to the PSF, we applied the correction factors of 1/$f_{50}$, to our flux measurements in each bin. 

\subsection{X-ray Count Distribution}

\begin{figure*}
\begin{center}
\setlength{\unitlength}{1mm}
\begin{picture}(150,60)
\put(-10,-10){\includegraphics{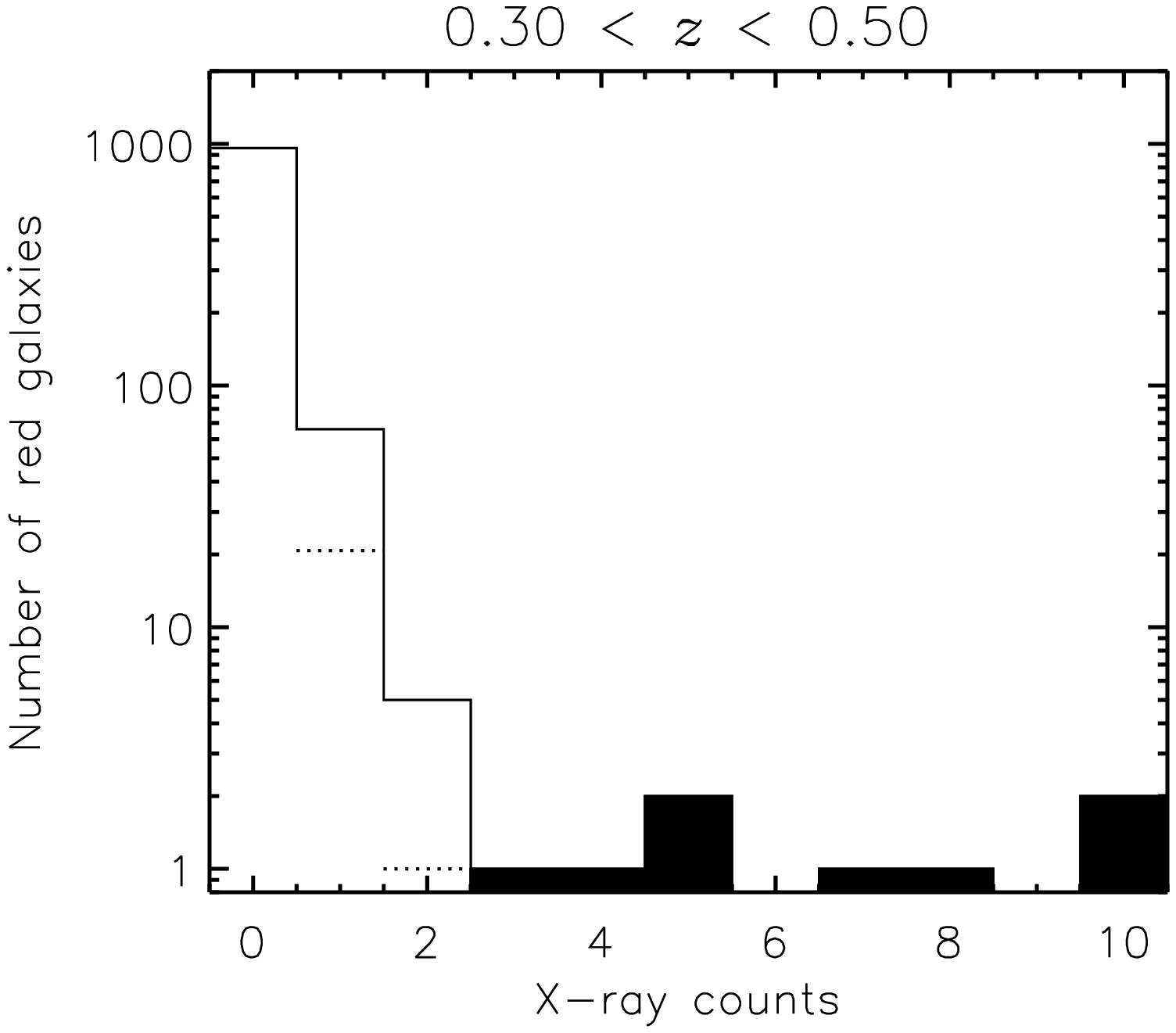}}
\put(45,-10){\includegraphics{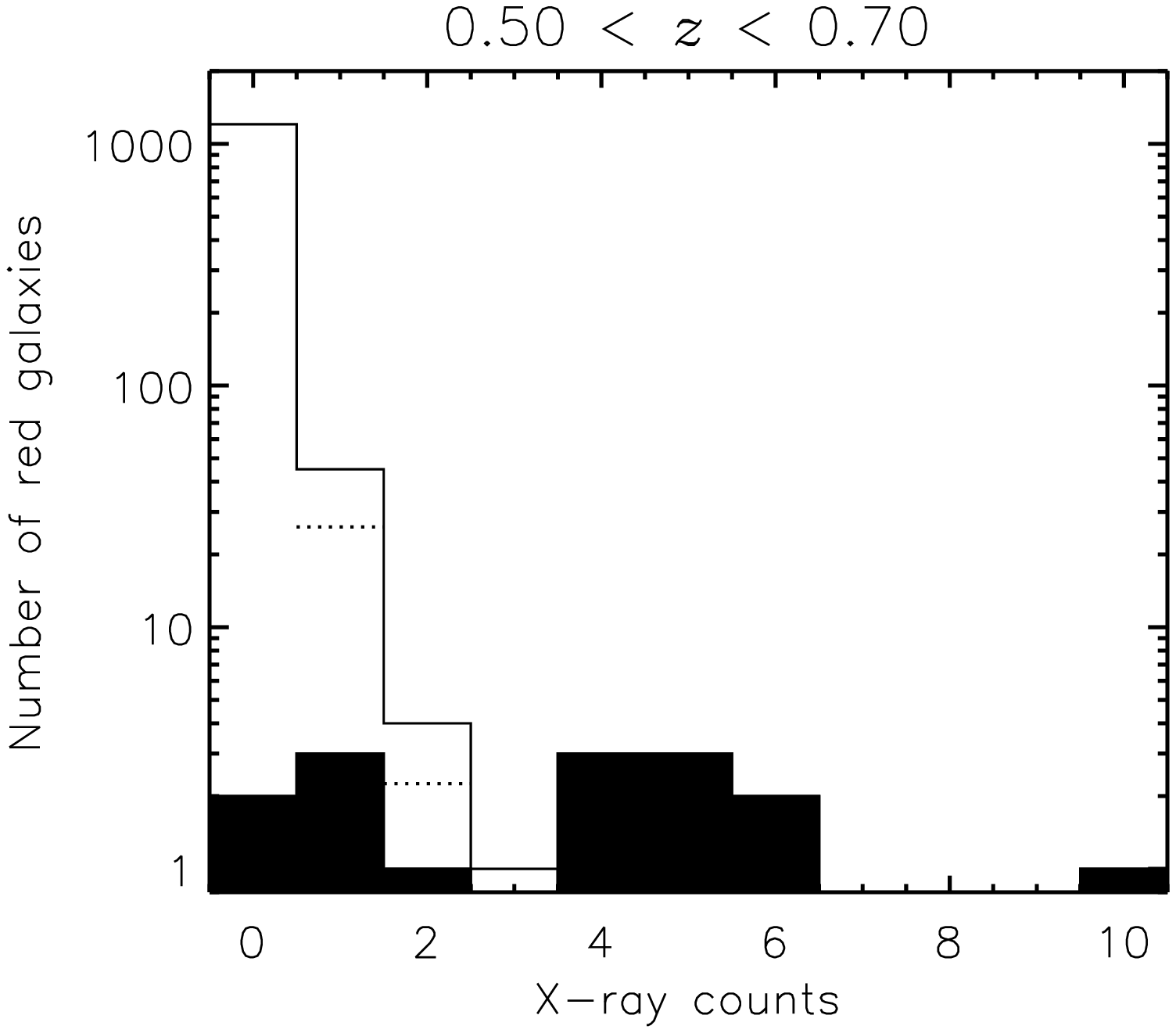}}
\put(100,-10){\includegraphics{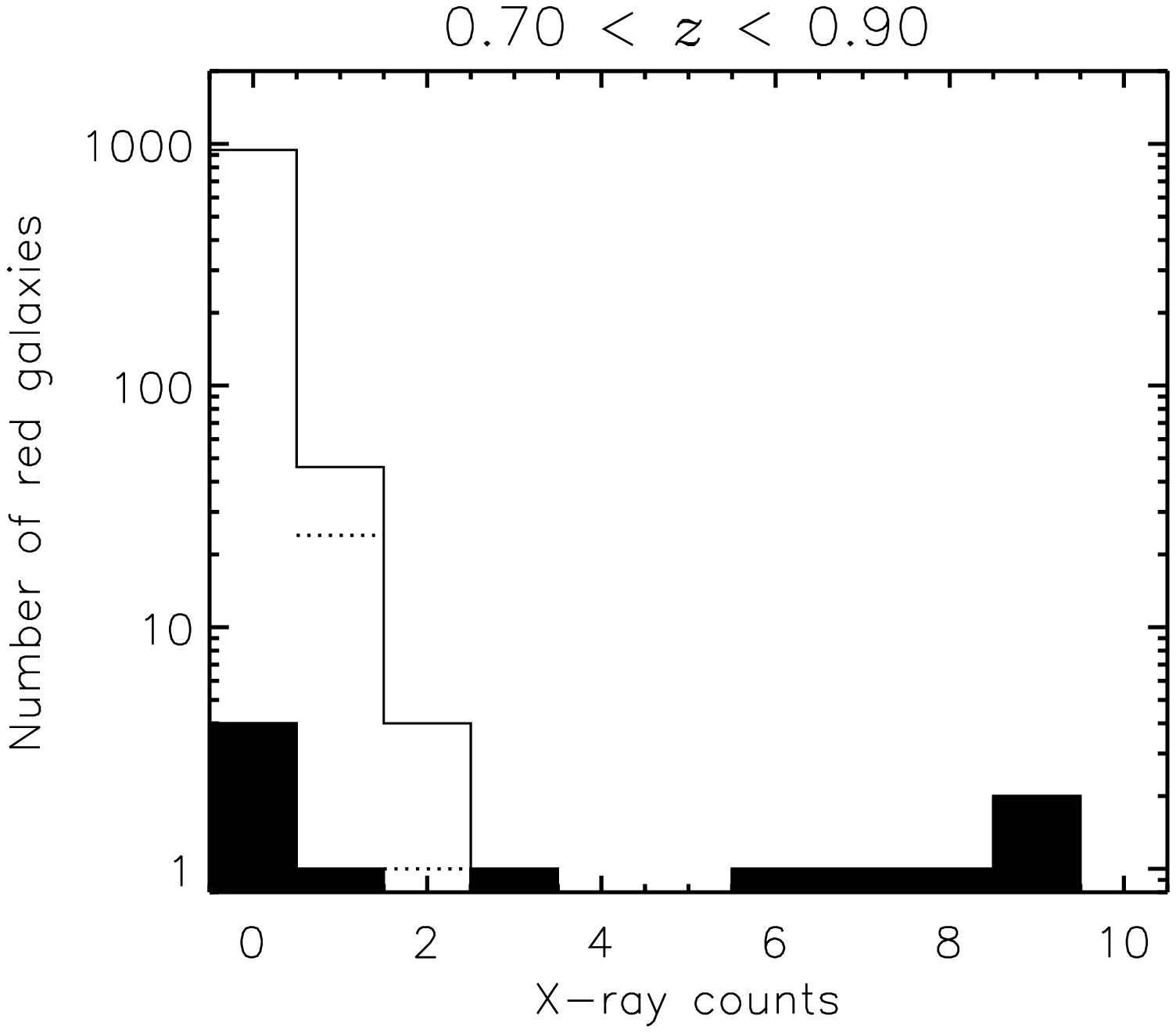}}
\end{picture}
\end{center}
{\caption[junk]{\label{fig:hist} Histograms showing the number of photons detected within the 1.5 arcsec aperture radius as a function of the number of red galaxies for the 0.3$<z<$0.5 (left), 0.5$<z<$0.7 (center), and 0.7$<z<$0.9 (right) redshift bins.  Filled bins show the distribution of the sources which are individually detected in the X-ray ($\ge$4 total counts using {\small WAVDETECT}). These sources may have less than 4 counts within the 1.5 arcsec aperture radius due to PSF effects. The 10 count bin includes all sources with $>$10 counts (there is one detected source in the 0.3$<z<$0.5 redshift bin with 24 X-ray counts). The dotted lines show the expected number of background counts. }}
\end{figure*}

Figure~\ref{fig:hist} shows the distribution of counts contributing to the stacked images and identifies the detected sources. After subtracting the individually detected sources, only $\sim$ 7\%, 4\%, and 5\% of the galaxies contribute to the counts in our stacked signal in the $0.3<z<0.5$, $0.5<z<0.7$, and $0.7<z<0.9$ redshift bins respectively. However, assuming all sources have the average PSF, there may be as many as 17\%, 11\%, and 13\% which would contribute to the signal if all their counts didn't fall outside the 1.5 arcsec aperture radius. What we cannot tell from the counts is the nature of the true luminosity distribution. The stacking analysis by construction assumes that the one- and two-count sources are merely the bright tail in the Poisson distribution of the X-ray flux, that the detected photons are drawn at random from the red galaxy population, and that the best estimate of the mean luminosity is derived by dividing the detected flux among all the stacked sources. This assumption is supported by the optical properties of the individually detected and undetected sources. We found no significant differences when comparing the distribution of optical absolute magnitudes and colors among the sources contributing one or more counts and the overall population. As far as we can tell, given the current sample, the individually detected and undetected sources appear to be drawn from the same optical population. If only a fraction of the red galaxies contributed to the stacked signal, then the mean flux of the X-ray-bright sub-population would be larger. Distributing the flux equally over the entire population, as we have done, therefore places a conservative lower bound on the mean flux.
 
To investigate the possible distribution of sources below our detection limit, some insight can be gained by considering the individually detected sources in the Chandra Deep Fields (\citealt{bra01}; \citealt{gia02}). \citet{szo03} present the spectroscopic follow-up of the X-ray sources detected in the 942-ks exposure of the Chandra Deep Field South (CDF-S). The CDF-S is a factor of $\approx$ 5 shallower than our stacked signal and $\approx$ 25 times smaller in area, but has the advantage that the properties of the individually detected sources can be distinguished. The CDF-S sample contains 15 X-ray-detected galaxies in the redshift range $0.3<z<0.9$ whose optical spectra show no obvious signs of AGN activity and resemble those of our red galaxy population. The combined hardness ratio of the CDF-S sources is 0.07, which suggests obscured AGN activity (e.g., \citealt{nor04}). This is consistent with the mean hardness ratios of $-$0.03, 0.15, and $-$0.07 for the red galaxies in the $0.3<z<0.5$, $0.5<z<0.7$, and $0.7<z<0.9$ redshift bins respectively. Assuming a power-law spectrum with photon index $\Gamma$=1.7, the mean luminosity in the 0.5-7 keV band of the CDF-S sources is $\approx 1 \times 10^{42}$ ergs s$^{-1}$. This corresponds to 0.3 counts in our 5-ks exposure. 
                                                                                           
All 15 detected CDF-S sources \citep{szo03} would fall below the 4-count X-ray detection limit of our XBo\"otes survey data. However, the count distribution of these 15 sources can be used to predict the number of 1, 2, and 3 photon count sources we would expect in our 5~ks observation. Assuming that our sample has exactly the same X-ray flux distribution as the 15 CDF-S sources, and correcting for our 1.5 arcsec aperture radius, we found that the number of sources anticipated from the CDF-S sample is similar to the number observed. The CDF-S data are therefore consistent with the hypothesis that the X-ray flux in our sample arises from a large fraction of the red galaxies. However, these numbers are presently very uncertain because of the small number of photons in our stacked data and the limited public studies available from the deep surveys. 

Analysis of the full (9 deg$^2$) survey will greatly increase our ability to study the predicted photon count distribution. When combined with the growing body of spectroscopic information from the Chandra deep surveys, these may place significant constraints on the underlying source distribution. We therefore defer a more detailed analysis to a future paper. Here, we simply state that the constraints we can currently place from the published deeper Chandra surveys are consistent with our assumption that the stacked signal is representative of the overall red galaxy population, and not simply the result of a small, X-ray luminous, outlier population.

\section{Redshift Evolution of the Mean X-ray Luminosity}

We can use our three sub-samples to investigate the evolution in the mean X-ray luminosity of the red galaxy population. Once we have determined this, we would like to test that no systematic effects are introducted by our stacking method. 

\subsection{The Observed Redshift Evolution}

In order to convert the X-ray counts to an X-ray flux, one must assume a model for the spectral energy distribution. When we assume a canonical AGN power-law spectrum with photon index $\Gamma$=1.7 and the Galactic HI column density of 1.75$\times$10$^{20}$ cm$^{-2}$ \citep{sta92}, we significantly underestimate the fraction of photons in the hard-band relative to the total-band (0.25 versus the observed fraction of 0.5). This is clearly inconsistent with our data and indicates that the shape of the assumed spectral energy distribution is incorrect. If we instead assume an intrinsic power-law spectrum with photon index $\Gamma$=1.7, the relative count rates in the hard and soft bins can be explained in a consistent scenario whereby the continuum is absorbed by an HI column density of $\approx 2 \times 10^{22}$ cm$^{-2}$ (which will absorb a much larger number of the soft photons) at the redshift of the galaxy in all redshift bins.  Alternatively, the relative count rates can be explained by an unabsorbed power-law spectrum with photon index $\Gamma$=0.7.  

\begin{figure*}
\begin{center}
\setlength{\unitlength}{1mm}
\begin{picture}(150,60)
\put(0,-15){\includegraphics{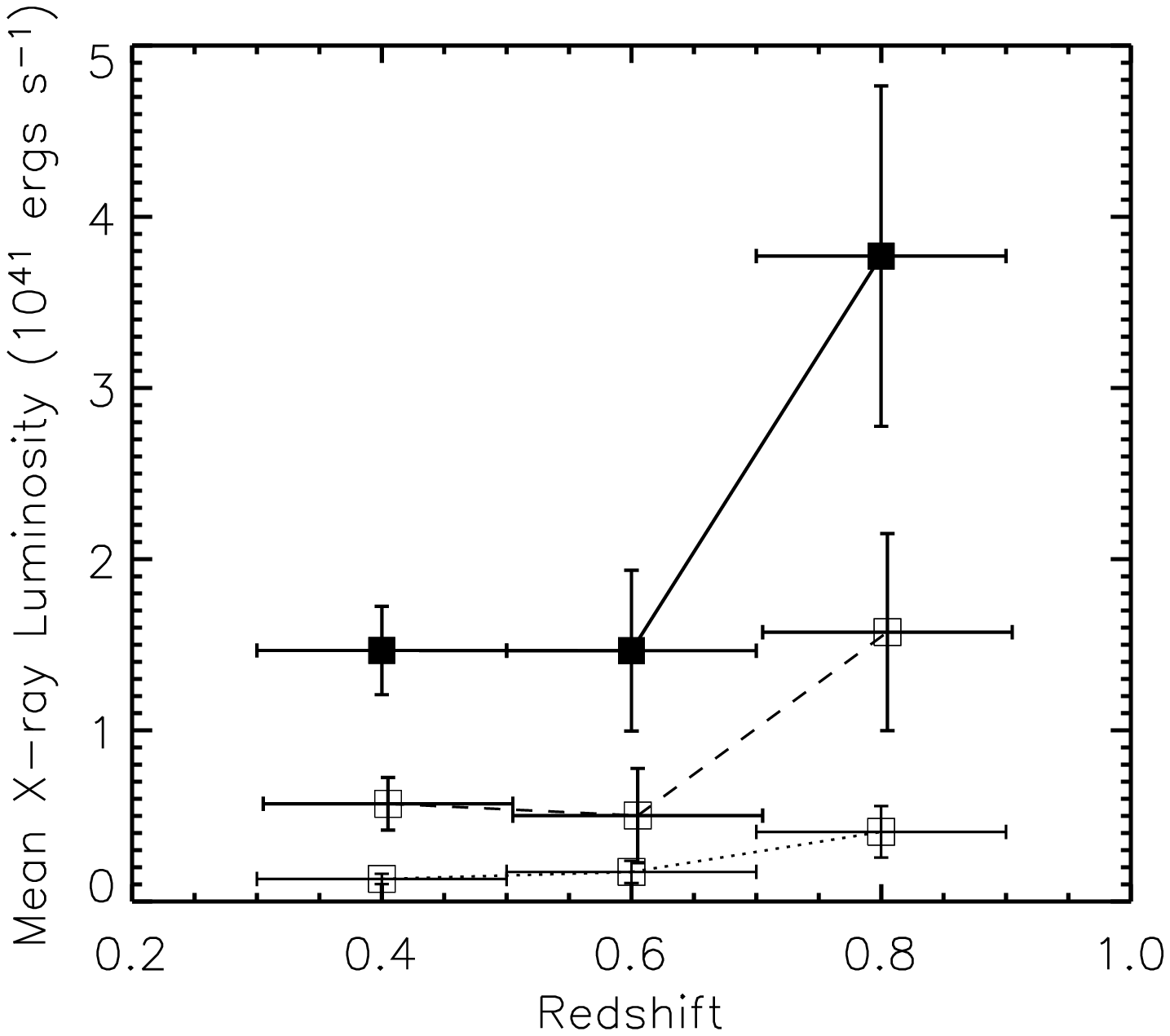}}
\put(70,-15){\includegraphics{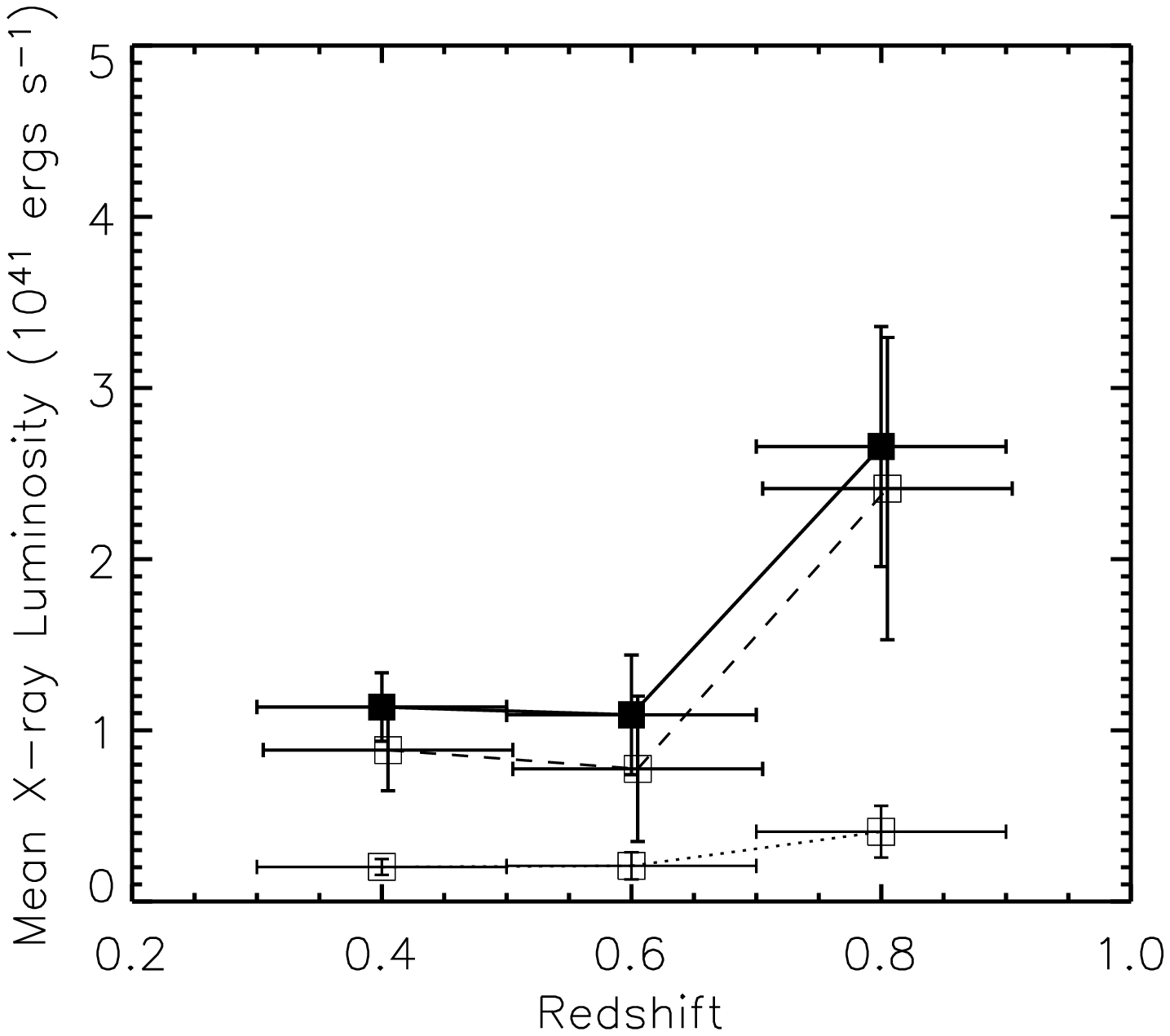}}
\end{picture}
\end{center}
{\caption[junk]{\label{fig:lumz} Redshift evolution of the mean X-ray luminosity (excluding individually X-ray detected sources) and the associated Poisson counting uncertainties for two different spectral energy distributions. The X-ray luminosity was calculated assuming a power-law spectrum with photon index $\Gamma$=1.7 and an additional HI column density of 2$\times$10$^{22}$ cm$^{-2}$ at the mean redshift of the bin (left) and with an alternative photon index, $\Gamma$=0.7 (right). In both cases, we assumed a Galactic HI column density of 1.75$\times$10$^{20}$ cm$^{-2}$. The total, hard, and soft luminosities are represented by the solid, dashed, and dotted lines respectively. The luminosities obtained when including the detected sources can be found in Table~\ref{tab:results}.
}}
\end{figure*}

Figure~\ref{fig:lumz} shows the evolution of the mean X-ray luminosity of the red galaxy population as a function of redshift in the total, soft, and hard X-ray bands, based on these two spectral models. Both the total and hard band luminosities increase significantly with redshift in both cases. Our conclusions are not strongly dependent on the assumed spectral model. If we include the contribution from the detected sources, the overall trend does not change (cf., Table~1).

\subsection{Monte Carlo Simulations}
\label{sec:mc}

In order to assess our ability to recover the true redshift evolution of the X-ray luminosity evolution using the stacking approach, we performed Monte Carlo simulations of our stacking analysis. We generated synthetic red galaxy catalogs distributed randomly within an ACIS field. We assumed a simple input model for the evolution of total X-ray luminosity with redshift ($L_X\propto(1+z)^3$), normalized to the observed mean luminosity in the $0.3<z<0.5$ redshift bin. This is a trend that we might expect from \citet{bar01}. The X-ray photons from each galaxy were distributed using a Gaussian PSF model that varies in its 50\% encircled energy radius as a function of off-axis distance according to Eqn.~\ref{eqn:r50}. We modeled the ACIS background by distributing photons at random positions in the field based on the observed background surface densities of 0.0032 total (0.5-7 keV) photons per square arcsec.

\begin{figure}[h]
\begin{center}
\setlength{\unitlength}{1mm}
\begin{picture}(40,60)
\put(-15,-15){\includegraphics{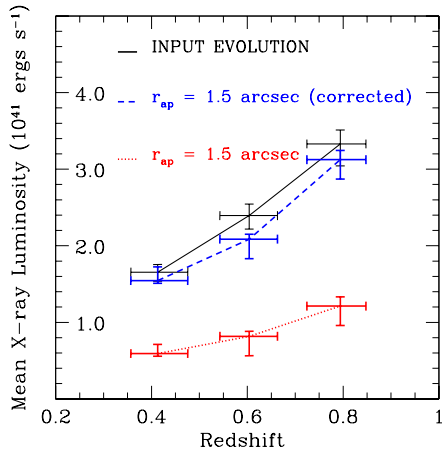}}
\end{picture}
\end{center}
{\caption[junk]{\label{fig:mc} The redshift evolution of the mean X-ray luminosity assumed in the Monte Carlo simulations (solid line) and recovered using using a 1.5 arcsec aperture radius both with (dashed line) and without (dotted line) applying the PSF corrections discussed in Section~3.1. The error bars in the plot were calculated using bootstrap resampling.}}
\end{figure}

Having simulated the galaxy and photon catalogs, we then followed the same stacking procedure that was employed for the real data. Figure~\ref{fig:mc} shows the assumed and recovered luminosity evolution based on a 1.5 arcsec radius aperture (r$_{\rm ap}$) and demonstrates that we successfully recover the input luminosity profile once the PSF correction factors (discussed in Section~3.1) have been applied. Although, adopting larger values of r$_{\rm ap}$ enabled us to recover a larger fraction of the signal, this introduced greater background contamination and correspondingly larger measurement errors. Because we can estimate the PSF corrections with reasonable accuracy, we conclude that the combined use of a smaller aperture and PSF corrections provides the best strategy for stacking the emission of point-source dominated galaxies, like those in our sample.


\section{Discussion}

\subsection{X-ray Emission Mechanisms}

The total X-ray emission from galaxies may contain contributions from a variety of sources: stellar objects, such as low-mass and high-mass X-ray binaries (hereinafter LMXBs and HMXBs); diffuse hot gas;  and AGN (e.g., \citealt{fab89}). The LMXB population is long-lived and generally traces the old stellar population in galaxies and hence the stellar mass of the galaxy; in contrast, the HMXBs have short lifetimes and the luminosity of this population reflects the star-formation rate in galaxies.

\subsubsection{Contribution of Stellar Sources}

We can estimate the integrated X-ray luminosity expected from LMXBs from relations derived from high-spatial-resolution X-ray observations of nearby galaxies. Adopting the relation given by \citet{kim04} between the absolute $K$ magnitude of early-type galaxies and their integrated LMXB X-ray luminosity, and assuming that this relation does not vary with redshift, we predict that LMXBs contribute a mean X-ray luminosity of ${\rm L_{LMXB}}\approx$0.2 $\times 10^{41}$ ergs s$^{-1}$ in all redshift bins (see Table~\ref{tab:results}). We note that using the relation of \citet{col04} results in luminosities a factor of $\approx$10 smaller but this is probably an underestimate because they only calculate the integrated luminosity of discrete sources. A comparison with Table~\ref{tab:results} and Figure~\ref{fig:lumz} illustrates that this contribution is small compared with the mean X-ray luminosity from the red galaxy sample; summing over all the galaxies, we find that LMXBs should contribute $\approx$ 11\%, 9\%, and 4\% of the stacked total luminosity in the $0.3<z<0.5$, $0.5<z<0.7$, and $0.7<z<0.9$ redshift bins respectively.

If the number of LMXBs within a galaxy is larger at $z=0.3-0.9$ than in the local Universe, their contributions to the total stacked luminosity could be higher. \citet{whi98} suggest that the X-ray luminosity of galaxies at $z\sim$0.5-1 should be an order of magnitude higher than in the local Universe due to the effects of the declining cosmic star formation rate on the evolution of the LMXB population. However, stacking analyses of optically bright HDF-N galaxies at $z\sim$0.5 show that the X-ray luminosities of these galaxies can only be up to a factor of $\sim$2 times higher than local galaxies with similar optical properties \citep{bra01c} and models have been revised accordingly (\citealt{gho01}; \citealt{pta01}). Including such an evolution suggests that the contribution of LMXBs to the total stacked luminosity is at most $\approx$ 20\%.

We can also estimate the contribution to the X-ray luminosity from HMXBs by scaling from the star-formation rate (SFR) estimates derived from the population synthesis model fits to the optical photometry. The mean SFRs estimated for the $0.3<z<0.5$, $0.5<z<0.7$, and $0.7<z<0.9$ redshift bins are $1.8{~\rm M_\odot\, yr^{-1}}$, $2.3{~\rm M_\odot\, yr^{-1}}$, and $4.1{~\rm M_\odot\, yr^{-1}}$. For each galaxy, we used the scaling of \citet{gri03} in either the high-SFR ($>4.5{~\rm M_\odot\, yr^{-1}}$; linear) or low-SFR ($<4.5{~\rm M_\odot\, yr^{-1}}$; non-linear) regime to estimate the hard X-ray luminosity from HMXBs. We calculate hard X-ray luminosities of ${\rm L_{HMXB}}\approx$ 0.10 $\times 10^{41}$ ergs s$^{-1}$, 0.14 $\times 10^{41}$ ergs s$^{-1}$, and 0.26  $\times 10^{41}$ ergs s$^{-1}$ originating from HMXBs: i.e., $\approx$ 7\%, 9\%, and 7\% of the stacked hard X-ray luminosity in each of the three redshift bins. 

There may also be a small contribution in the X-ray emission arising from hot gas that typically surrounds elliptical galaxies \citep{for94}. This emission is predominantly in the soft band and thus subtracting any contamination would make our underlying X-ray spectrum even harder. We note, however, that a comparison of the radial profiles with the PSF shown in Figure~\ref{fig:psf_data} show no compelling evidence for a significant extended component.

We conclude that the large mean X-ray luminosity ($\sim$10$^{41}$ ergs s$^{-1}$) and hard spectrum of the red galaxies strongly suggests that the signal must be dominated by X-ray emission from accretion onto a central AGN rather than X-ray emission originating from stellar sources.

\subsubsection{AGN Accretion Processes}

We have shown in Section~4.1 that the mean red galaxy hardness ratio can be reproduced by either an absorbed (${\rm N_H\approx 2 \times 10^{22}\,cm^{-2}}$) $\Gamma$=1.7 power-law source or an unabsorbed source with a harder ($\Gamma$=0.7) spectrum.

An absorbed (${\rm N_H\approx 2 \times 10^{22}\,cm^{-2}}$) power-law spectrum with a canonical slope, $\Gamma$=1.7 is consistent with that of a population of moderately obscured, Seyfert-like AGN (\citealt{nan94}; \citealt{ris99}), whose primary emission mechanism is thought to be inverse-Compton scattering originating from a hot accretion disk. The optical and soft X-ray photons are thought to be absorbed by an intervening column of gas and dust resulting in a hard X-ray spectrum and no optical AGN signature. \citet{ued03} find that X-ray absorbed AGNs make up a larger fraction of fainter AGN populations, suggesting that absorption may indeed be significant in our sample.


A harder ($\Gamma$=0.7) spectrum could be produced by accretion onto a SMBH via a radiatively inefficient accretion flow, such as an advection-dominated accretion flow (ADAF), which has a harder characteristic X-ray spectrum than other mechanisms and thus requires little or no obscuration to account for the hard spectrum. ADAFs were originally proposed by \citet{ich77} and \citet{ree82}, and have been invoked to explain the lack of strong X-ray emission expected from accretion of the extensive hot gas known to exist in elliptical galaxies (e.g., \citealt{nar95}; \citealt{qua99}; \citealt{bla04}). \citet{all00} and \citet{dim00} analyzed the X-ray spectra of 6 nearby giant elliptical galaxies observed with ASCA and found a best-fit photon index of $\Gamma$=0.6-1.5 and suggest that this is best explained by a ADAF. However, higher resolution imaging by Chandra has shown that much of the hard X-ray emission spatially unresolved by ASCA originates from luminous LMXBs (\citealt{ang01}; \citealt{loe01}). 

Although the true underlying X-ray luminosity distribution of the red galaxy population is unclear, it appears likely that the majority of the population emits a hard X-ray spectrum at a luminosity that is too high to be due to stellar sources and is instead powered by accretion onto a SMBH. The results of several studies suggest that only half of all local galaxies have evidence for nuclear point sources and/or exhibit AGN signatures in their optical spectra (e.g., \citealt{ho97}; \citealt{sar99}; \citealt{rav01}). However, this fraction is uncertain due to difficulties in resolving nuclear point sources, distinguishing them from compact nuclear star clusters, and identifying AGN in spectra dominated by the host galaxy light. In addition, if the accretion flow has become advection dominated, the radiative signature is likely to be too weak to detect at optical wavelengths. 

\subsection{Contribution to the Hard X-ray Background}
\label{sec:bkgd}

If obscured AGN and/or low radiative efficiency accretion flows onto SMBHs exist in most early type galaxies, their integrated emission may contribute significantly to the hard X-ray background \citep{dim99}. 

Approximately 94\% and 89\% of the soft and hard cosmic X-ray background (CXB) respectively has now been resolved into discrete sources \citep{mor03}. The CXB spectrum is well described by a photon index of $\Gamma\sim$1.4 (e.g., \citealt{mus00}). To fit the shape of the CXB, models require an as yet unresolved population of hard spectrum sources; typically, a population of heavily obscured AGN is invoked (e.g., \citealt{ued03}). 

The total hard X-ray flux of our red galaxy population (including X-ray detected sources) is $\sim$ 1 $\times 10^{-12}$ ergs s$^{-1}$ cm$^{-2}$ deg$^{-2}$. This sample therefore contributes approximately 5\% of the hard CXB. This is a lower limit as it only includes a sub-sample of normal galaxies optically dominated by stellar light within the redshift range 0.3$< z <$0.9. More luminous AGN within similar galaxies will have been excluded from our analysis because they significantly modify the optical colors of their hosts and cause them to drop out of our sample. The fraction of the CXB contributed by all bulge dominated galaxies is therefore expected to be much higher.    

The majority of our red galaxy sample would be detected at the depth of the Chandra Deep fields. However, if a large fraction of normal galaxies harbor very low luminosity AGN (i.e., below the limit of the deep surveys), they could contribute non-negligibly to the unresolved portion of the hard CXB, because of their high number density.

\subsection{The Nuclear Accretion History of Red Galaxies}

Our stacking analysis of the red galaxy population has demonstrated that their mean X-ray luminosity decreases with time. This is perhaps as expected: the dramatic fall in the space density of powerful AGN is well known, and has been attributed to a decrease in the fuel supply, fueling rate, and/or radiative efficiency onto the SMBH (e.g., \citealt{cav00}). We observe a redshift evolution in the total X-ray luminosity of $(1+z)^{4.0\pm2.4}$. This is consistent with the $\sim (1+z)^3$ variation in accretion activity inferred by \citet{bar01} from X-ray sources in the CDF-N, and, in corroboration with the above observations, is indicative of a significant decline in the mean rate of accretion onto SMBHs in early-type galaxies from $z \sim 1$ to the present. 


However, we need to be careful when comparing our results to those derived from X-ray selected samples. Recent results show that the peak in the number density of luminous AGN occurs at higher redshift than that of the less powerful AGN (\citealt{cow03}; \citealt{ued03}): the so-called ``cosmic down-sizing'' effect. This suggests that at lower redshift, progressively smaller galaxies are contributing a larger fraction of the total X-ray luminosity. Because X-ray selected samples trace the evolution of the overall X-ray luminosity, regardless of whether the contribution from different galaxy populations changes, the redshift evolution of an X-ray selected sample may be less dramatic than a sample which traces the evolution in the X-ray luminosity of a particular galaxy population. 

By selecting our red galaxies to have the same evolution-corrected, absolute $R$-band magnitudes, we are tracing the evolution in the X-ray properties of the same galaxy population as a function of redshift. We are therefore measuring an intrinsic drop in accretion activity of this galaxy population. As discussed in Section~\ref{sec:bkgd}, more luminous unobscured AGN within similar host galaxies will have been excluded from our analysis due to the modification in the optical color of the host galaxy due to the AGN. The redshift evolution of the underlying population is therefore likely to be even stronger than is observed. Since the overwhelming majority of luminous AGN at high redshift are hosted by luminous early-type galaxies (e.g., \citealt{mcl99}; \citealt{kau03}), we are likely to be be tracing the same population at lower redshifts. The decrease in X-ray luminosity with time in our sample therefore suggests we are witnessing the tailing off of the accretion activity onto SMBHs in early-type galaxies since the more powerful quasar epoch. 

\section{Conclusions}

Our primary conclusions are as follows:

\noindent\textbullet ~We have detected significant X-ray emission from the red galaxy population in the NDWFS in each of the $0.3<z<0.5$, $0.5<z<0.7$, and $0.7<z<0.9$ redshift bins. \\
\textbullet ~The large X-ray luminosity and hard spectral slope strongly suggest that the dominant X-ray emission mechanism is accretion onto a SMBH. The spectral energy distribution can be reproduced by invoking either a moderately obscured AGN (${\rm N_H\approx 2 \times 10^{22}\,cm^{-2}}, \Gamma$=1.7) or a harder ADAF-like ($\Gamma$=0.7) spectrum.\\
\textbullet ~The mean hard X-ray luminosity of the red galaxy population increases as a function of redshift. Monte Carlo simulations show that our method will correctly recover any intrinsic evolution. \\
\textbullet ~The hard X-ray emission produced by this population contributes at least 5\% of the hard X-ray background. \\
\textbullet ~We interpret the redshift evolution of the X-ray luminosity as a global decline in the mean AGN activity of normal early-type galaxies.\\

One intriguing aspect of our observations is the relative hardness of the mean X-ray spectrum in the stacked data. We have interpreted this as resulting from either an absorbed AGN spectrum or a radiatively inefficient accretion flow (such as an ADAF). In the latter case, we envision a scenario where the accretion onto SMBHs turns progressively more radiatively inefficient, resulting in a harder mean spectrum and lower X-ray luminosity with decreasing redshift. Our results suggest that nuclear accretion is still occurring in most present-day early-type galaxies, but with lower efficiency and/or in a more heavily obscured regime.

To distinguish between these possibilities and estimate the true accretion rate, we intend to extend this study to encompass the full 9 deg$^{2}$ survey area. The larger sample will contain an order of magnitude more galaxies. It will therefore enable us to determine the best-fit model to the X-ray spectral energy distribution and hence quantify the typical accretion rate onto SMBHs in bulge dominated galaxies and its evolution with redshift.

\acknowledgements{We thank the NDWFS survey team, particularly Jenna Claver, Alyson Ford, Lissa Miller, and Glenn Tiede for assistance in providing the optical data in this work. We thank the anonymous referee for his/her useful comments. Our research is supported by the National Optical Astronomy Observatory which is operated by the Association of Universities for Research in Astronomy, Inc. (AURA) under a cooperative agreement with the National Science Foundation. Support for this work was provided by the National Aeronautics and Space Administration through Chandra Award Number GO3-4176 issued by the Chandra X-ray Observatory Center, which is operated by the Smithsonian Astrophysical Observatory for and on behalf of the National Aeronautics and Space Administration under contract NAS8-39073.}

\bibliography{ms}

\begin{thebibliography}{56}
\expandafter\ifx\csname natexlab\endcsname\relax\def\natexlab#1{#1}\fi

\bibitem[{{Allen} {et~al.}(2000){Allen}, {Di Matteo}, \& {Fabian}}]{all00}
{Allen}, S.~W., {Di Matteo}, T., \& {Fabian}, A.~C. 2000, \mnras, 311, 493

\bibitem[Angelini et al.(2001)]{ang01} Angelini, L., 
Loewenstein, M., \& Mushotzky, R.~F.\ 2001, \apjl, 557, L35 

\bibitem[{{Barger} {et~al.}(2001){Barger}, {Cowie}, {Bautz}, {Brandt},
  {Garmire}, {Hornschemeier}, {Ivison}, \& {Owen}}]{bar01}
{Barger}, A.~J., {Cowie}, L.~L., {Bautz}, M.~W., {Brandt}, W.~N., {Garmire},
  G.~P., {Hornschemeier}, A.~E., {Ivison}, R.~J., \& {Owen}, F.~N. 2001, \aj,
  122, 2177

\bibitem[{{Begelman}(2003)}]{beg03}
{Begelman}, M.~C. 2003, Science, 300, 1898

\bibitem[{{Bertin} \& {Arnouts}(1996)}]{ber96}
{Bertin}, E., \& {Arnouts}, S. 1996, \aaps, 117, 393

\bibitem[{{Blandford} \& {Begelman}(2004)}]{bla04}
{Blandford}, R.~D., \& {Begelman}, M.~C. 2004, \mnras, 349, 68

\bibitem[{{Brandt} {et~al.}(2001{\natexlab{a}}){Brandt} {et~al.}}]{bra01}
{Brandt}, W.~N., {et~al.}, 2001{\natexlab{a}}, \aj, 122, 2810

\bibitem[{{Brandt} {et~al.}(2001{\natexlab{b}}){Brandt}, {Hornschemeier},
  {Schneider}, {Alexander}, {Bauer}, {Garmire}, \& {Vignali}}]{bra01b}
{Brandt}, W.~N., {Hornschemeier}, A.~E., {Schneider}, D.~P., {Alexander},
  D.~M., {Bauer}, F.~E., {Garmire}, G.~P., \& {Vignali}, C. 2001{\natexlab{b}},
  \apjl, 558, L5

\bibitem[Brandt et al.(2001{\natexlab{c}})]{bra01c} Brandt, W.~N., et al.\ 
2001{\natexlab{c}}, \aj, 122, 1

\bibitem[{{Brown} {et~al.}(2003){Brown}, {Dey}, {Jannuzi}, {Lauer}, {Tiede}, \&
  {Mikles}}]{bro03}
{Brown}, M.~J.~I., {Dey}, A., {Jannuzi}, B.~T., {Lauer}, T.~R., {Tiede}, G.~P.,
  \& {Mikles}, V.~J. 2003, \apj, 597, 225

\bibitem[{{Cavaliere} \& {Vittorini}(2000)}]{cav00}
{Cavaliere}, A., \& {Vittorini}, V. 2000, \apj, 543, 599


\bibitem[{{Colbert} {et~al.}(2004){Colbert}, {Heckman}, {Ptak}, {Strickland},
  \& {Weaver}}]{col04}
{Colbert}, E.~J.~M., {Heckman}, T.~M., {Ptak}, A.~F., {Strickland}, D.~K., \&
  {Weaver}, K.~A. 2004, \apj, 602, 231

\bibitem[{{Cowie} {et~al.}(2003){Cowie}, {Barger}, {Bautz}, {Brandt}, \&
  {Garmire}}]{cow03}
{Cowie}, L.~L., {Barger}, A.~J., {Bautz}, M.~W., {Brandt}, W.~N., \& {Garmire},
  G.~P. 2003, \apjl, 584, L57

\bibitem[{{Csabai} {et~al.}(2003)}]{csa03}
{Csabai}, I., {et~al.}, 2003, \aj, 125, 580

\bibitem[{{di Matteo} \& {Allen}(1999)}]{dim99}
{di Matteo}, T., \& {Allen}, S.~W. 1999, \apjl, 527, L21

\bibitem[{{di Matteo} {et~al.}(2000){di Matteo}, {Quataert}, {Allen},
  {Narayan}, \& {Fabian}}]{dim00}
{Di Matteo}, T., {Quataert}, E., {Allen}, S.~W., {Narayan}, R., \& {Fabian},
  A.~C. 2000, \mnras, 311, 507

\bibitem[{{Dunlop} {et~al.}(2003){Dunlop}, {McLure}, {Kukula}, {Baum}, {O'Dea},
  \& {Hughes}}]{dun03}
{Dunlop}, J.~S., {McLure}, R.~J., {Kukula}, M.~J., {Baum}, S.~A., {O'Dea},
  C.~P., \& {Hughes}, D.~H. 2003, \mnras, 340, 1095

\bibitem[{{Fabbiano}(1989)}]{fab89}
{Fabbiano}, G. 1989, \araa, 27, 87

\bibitem[{{Ferrarese} \& {Merritt}(2000)}]{fer00}
{Ferrarese}, L., \& {Merritt}, D. 2000, \apjl, 539, L9

\bibitem[{{Fioc} \& {Rocca-Volmerange}(1997)}]{fio97}
{Fioc}, M., \& {Rocca-Volmerange}, B. 1997, \aap, 326, 950

\bibitem[{{Forman} {et~al.}(1994){Forman}, {Jones}, \& {Tucker}}]{for94}
{Forman}, W., {Jones}, C., \& {Tucker}, W. 1994, \apj, 429, 77

\bibitem[{{Fukugita} {et~al.}(1995){Fukugita}, {Shimasaku}, \&
  {Ichikawa}}]{fuk95}
{Fukugita}, M., {Shimasaku}, K., \& {Ichikawa}, T. 1995, \pasp, 107, 945

\bibitem[{{Gebhardt} {et~al.}(2000)}]{geb00}
{Gebhardt}, K.,{et~al.}, 2000, \apjl, 539, L13

\bibitem[{{Giacconi} {et~al.}(2002)}]{gia02}
{Giacconi}, R.,{et~al.}, 2002, \apjs, 139, 369

\bibitem[Ghosh \& White(2001)]{gho01} Ghosh, P., \& White, 
N.~E.\ 2001, \apjl, 559, L97

\bibitem[{{Grimm} {et~al.}(2003){Grimm}, {Gilfanov}, \& {Sunyaev}}]{gri03}
{Grimm}, H.-J., {Gilfanov}, M., \& {Sunyaev}, R. 2003, \mnras, 339, 793

\bibitem[{{Grupe} \& {Mathur}(2004)}]{gru04}
{Grupe}, D., \& {Mathur}, S. 2004, \apjl, 606, L41

\bibitem[Heckman et al.(2004)]{hec04} Heckman, T.~M., 
Kauffmann, G., Brinchmann, J., Charlot, S., Tremonti, C., \& White, 
S.~D.~M.\ 2004, \apj, 613, 109

\bibitem[{{Ho} {et~al.}(1997){Ho}, {Filippenko}, \& {Sargent}}]{ho97}
{Ho}, L.~C., {Filippenko}, A.~V., \& {Sargent}, W.~L.~W. 1997, \apjs, 112, 315

\bibitem[{{Hogg} {et~al.}(2002)}]{hog02}
{Hogg}, D.~W., {et~al.}, 2002, \aj, 124, 646

\bibitem[{{Ichimaru}(1977)}]{ich77}
{Ichimaru}, S. 1977, \apj, 214, 840

\bibitem[{{Im} {et~al.}(2002)}]{im02}
{Im}, M., {et~al.}, 2002, \apj, 571, 136

\bibitem[{{Jannuzi} \& {Dey}(1999)}]{jan99}
{Jannuzi}, B.~T., \& {Dey}, A. 1999, in "Photometric Redshifts and the
  Detection of High Redshift Galaxies", ASP Conference Series, Vol. 191, Edited
  by R. Weymann, L. Storrie-Lombardi, M. Sawicki, and R. Brunner., 111

\bibitem[{{J{\o}rgensen} {et~al.}(1999){J{\o}rgensen}, {Franx}, {Hjorth}, \&
  {van Dokkum}}]{jor99}
{J{\o}rgensen}, I., {Franx}, M., {Hjorth}, J., \& {van Dokkum}, P.~G. 1999,
  \mnras, 308, 833

\bibitem[{{Kauffmann} {et~al.}(2003)}]{kau03}
{Kauffmann}, G., {et~al.}, 2003, \mnras, 346, 1055

\bibitem[{{Kim} {et~al.}(2004)}]{kim04}
{Kim}, D.~W., {et~al.}, 2004, \apjs, 150, 19

\bibitem[{{Kochanek} {et~al.}(2000)}]{koc00}
{Kochanek}, C.~S., {et~al.}, 2000, \apj, 543, 131

\bibitem[{{Kormendy} \& {Richstone}(1995)}]{kor95}
{Kormendy}, J., \& {Richstone}, D. 1995, \araa, 33, 581

\bibitem[Loewenstein et al.(2001)]{loe01} Loewenstein, M., 
Mushotzky, R.~F., Angelini, L., Arnaud, K.~A., \& Quataert, E.\ 2001, 
\apjl, 555, L21 

\bibitem[{{Magorrian} {et~al.}(1998)}]{mag98}
{Magorrian}, J., {et~al.}, 1998, \aj, 115, 2285

\bibitem[{{McLeod} \& {McLeod}(2001)}]{mcle01}
{McLeod}, K.~K., \& {McLeod}, B.~A. 2001, \apj, 546, 782

\bibitem[{{McLure} {et~al.}(1999){McLure}, {Kukula}, {Dunlop}, {Baum}, {O'Dea},
  \& {Hughes}}]{mcl99}
{McLure}, R.~J., {Kukula}, M.~J., {Dunlop}, J.~S., {Baum}, S.~A., {O'Dea},
  C.~P., \& {Hughes}, D.~H. 1999, \mnras, 308, 377

\bibitem[{{Moretti} {et~al.}(2003){Moretti}, {Campana}, {Lazzati}, \&
  {Tagliaferri}}]{mor03}
{Moretti}, A., {Campana}, S., {Lazzati}, D., \& {Tagliaferri}, G. 2003, \apj,
  588, 696

\bibitem[{{Mushotzky} {et~al.}(2000){Mushotzky}, {Cowie}, {Barger}, \&
  {Arnaud}}]{mus00}
{Mushotzky}, R.~F., {Cowie}, L.~L., {Barger}, A.~J., \& {Arnaud}, K.~A. 2000,
  \nat, 404, 459

\bibitem[{{Nandra} {et~al.}(2002){Nandra}, {Mushotzky}, {Arnaud}, {Steidel},
  {Adelberger}, {Gardner}, {Teplitz}, \& {Windhorst}}]{nan02}
{Nandra}, K., {Mushotzky}, R.~F., {Arnaud}, K., {Steidel}, C.~C., {Adelberger},
  K.~L., {Gardner}, J.~P., {Teplitz}, H.~I., \& {Windhorst}, R.~A. 2002, \apj,
  576, 625

\bibitem[{{Nandra} \& {Pounds}(1994)}]{nan94}
{Nandra}, K., \& {Pounds}, K.~A. 1994, \mnras, 268, 405

\bibitem[{{Narayan} \& {Yi}(1995)}]{nar95}
{Narayan}, R., \& {Yi}, I. 1995, \apj, 452, 710

\bibitem[{{Norman} {et~al.}(2004)}]{nor04}
{Norman}, C., {et~al.}, 2004, \apj, 607, 721

\bibitem[Ptak et al.(2001)]{pta01} Ptak, A., Griffiths, R., 
White, N., \& Ghosh, P.\ 2001, \apjl, 559, L91 

\bibitem[{{Quataert} \& {Narayan}(1999)}]{qua99}
{Quataert}, E., \& {Narayan}, R. 1999, \apj, 520, 298

\bibitem[{{Ravindranath} {et~al.}(2001){Ravindranath}, {Ho}, {Peng},
  {Filippenko}, \& {Sargent}}]{rav01}
{Ravindranath}, S., {Ho}, L.~C., {Peng}, C.~Y., {Filippenko}, A.~V., \&
  {Sargent}, W.~L.~W. 2001, \aj, 122, 653

\bibitem[{{Rees} {et~al.}(1982){Rees}, {Phinney}, {Begelman}, \&
  {Blandford}}]{ree82}
{Rees}, M.~J., {Phinney}, E.~S., {Begelman}, M.~C., \& {Blandford}, R.~D. 1982,
  \nat, 295, 17

\bibitem[{{Risaliti} {et~al.}(1999){Risaliti}, {Maiolino}, \&
  {Salvati}}]{ris99}
{Risaliti}, G., {Maiolino}, R., \& {Salvati}, M. 1999, \apj, 522, 157

\bibitem[Rusin et al.(2003)]{rus03} Rusin, D., et al.\ 2003, 
\apj, 587, 143 

\bibitem[{{Sarajedini} {et~al.}(1999){Sarajedini}, {Green}, {Griffiths}, \&
  {Ratnatunga}}]{sar99}
{Sarajedini}, V.~L., {Green}, R.~F., {Griffiths}, R.~E., \& {Ratnatunga}, K.
  1999, \apj, 514, 746

\bibitem[{{Schade} {et~al.}(1999)}]{sch99}
{Schade}, D., {et~al.}, 1999, \apj, 525, 31

\bibitem[{{Stanford} {et~al.}(1998){Stanford}, {Eisenhardt}, \&
  {Dickinson}}]{sta98}
{Stanford}, S.~A., {Eisenhardt}, P.~R., \& {Dickinson}, M. 1998, \apj, 492, 461

\bibitem[{{Stark} {et~al.}(1992){Stark}, {Gammie}, {Wilson}, {Bally}, {Linke},
  {Heiles}, \& {Hurwitz}}]{sta92}
{Stark}, A.~A., {Gammie}, C.~F., {Wilson}, R.~W., {Bally}, J., {Linke}, R.~A.,
  {Heiles}, C., \& {Hurwitz}, M. 1992, \apjs, 79, 77

\bibitem[Szokoly et al.(2004)]{szo03} Szokoly, G.~P., et al.\ 
2004, \apjs, 155, 271 

\bibitem[{{Ueda} {et~al.}(2003){Ueda}, {Akiyama}, {Ohta}, \& {Miyaji}}]{ued03}
{Ueda}, Y., {Akiyama}, M., {Ohta}, K., \& {Miyaji}, T. 2003, \apj, 598, 886


\bibitem[White \& Ghosh(1998)]{whi98} White, N.~E., \& Ghosh, 
P.\ 1998, \apjl, 504, L31 

\end{thebibliography}


\end{document}